\documentclass[twocolumn,amsmath,amssymb]{revtex4}
\usepackage{graphicx}
\usepackage{dcolumn}
\usepackage{bm}
\usepackage{relsize}
\usepackage{amsmath}
\usepackage{slashed}
\usepackage{amssymb}
\usepackage[mathscr]{eucal}
\usepackage{esdiff}
\usepackage{amsmath,color}
\usepackage{microtype}
\usepackage{verbatim} 
\DeclareMathAlphabet{\mathpzc}{OT1}{pzc}{m}{it}

\begin{document}
\def\al{\alpha}
\def\b{\beta}
\def\up{\upsilon}
\def\ni{\noindent} 
\def\th{\theta}
\def\lam{\lambda}
\def\Lam{\Lambda}
\def\sg{\sigma}
\def\O{\Omega}
\def\om{\omega}
\def\D{\Delta}
\def\de{\delta}
\def\ep{\epsilon}
\def\vphi{\varphi}
\def\up{\upsilon}
\def\t{\theta}
\def\chic#1{{\scriptscriptstyle #1}}
\def\xy{\rm{\chic{12}}}
\def\xz{\rm{\chic{13}}}
\def\be{\begin{equation}}
\def\ee{\end{equation}}
\def\bea{\begin{eqnarray}} 
\def \eea{\end{eqnarray}} 
\def\ba{\begin{array}}
\def\ea{\end{array}}
\def\ni{\noindent}
\def\hs{\hspace}
\def\vs{\vspace}
\def\hsp{\hspace{.03cm}}
\def\hsn{\hspace{-0.03cm}}
\def\ben{\begin{enumerate}}
\def\een{\end{enumerate}}
\def\bei{\begin{itemize}}
\def\eni{\end{itemize}}
\def\non{\nonumber}
\def\mb{\mathbf}
\def\mc{\mathcal}
\def\ms{\mathscr}
\def\mtt{\mathtt}
\def\mrm{\mathrm}
\def\mf{\mathfrak}
\def \Oxy{{\cal O}_{12}}
\def \Oxz{{\cal O}_{13}}
\def \Oyz{{\cal O}_{23}}
\def \G{\rm\Gamma}
\def \Hcal {{\cal H}}
\def\non{\nonumber}
\def\mtt{\mathtt}
\def\lan{\langle}
\def\ran{\rangle}
\def\dag{\dagger}
%

\newcommand{\dis}{\displaystyle}
\newcommand{\alfad}{\frac{\dis \bar \alpha_s}{\dis \pi}}
%
\newcommand{\fer}[1]{\textcolor[rgb]{0.627450980392157,0.125490196078431,0.941176470588235}{#1}}
\preprint{DRAFT 3.2}
%
\title{
Efficient numerical integration of neutrino oscillations in matter
}

\author{F.\;Casas}
\affiliation{Departament de Matem\`{a}tiques and IMAC, Universitat Jaume I,
  E-12071 Castell\'{o}n, Spain}
    \email{casas@mat.uji.es}

\author{J.\,C.\;D'Olivo} 
\affiliation{Instituto de Ciencias Nucleares, Universidad Nacional Aut\'onoma de 
M\'exico, Apartado Postal 70-543, 04510 M\'exico, D.F., Mexico}
  \email{dolivo@nucleares.unam.mx}
 
\author{J.\,A.\;Oteo}
\affiliation{Departamento de F\'{\i}sica Te\'orica, Universidad de
Valencia,  46100-Burjassot, Valencia, Spain}
  \email{oteo@uv.es}
  
\date{\today}

\begin{abstract}

A special purpose solver, based on the Magnus expansion, well suited for the integration of the linear 
three neutrino oscillations equations in matter is proposed. The computations are speeded up to two 
orders of magnitude with respect to a general numerical integrator, a fact that could smooth the way for 
massive numerical integration concomitant with experimental data analyses. Detailed illustrations about 
 numerical procedure and computer time costs are provided.

\end{abstract}

\maketitle

\section{Introduction}
\label{sec:intro}

Discovered with atmospheric \cite{kajita} and solar neutrinos \cite{mcdonald}, 
neutrino oscillations \cite{pontecorvo, mns} have been corroborated by experiments 
using neutrinos from nuclear reactors and accelerators \cite{zito}. 
These observations have established beyond doubt that neutrinos have masses and mix
\cite{petcov}, which has a fundamental impact not only on particle physics, but also on 
astrophysics and cosmology. For three flavors ($\nu_e,\nu_\mu,\nu_\tau$) the oscillation 
frequencies in vacuum are characterized by two independent differences between the 
squared masses:  $\de m^2 \equiv m^2_2 - m^2_1$  (solar) and 
$\D m^2 \equiv m^2_3 - (m^2_1 + m^2_2)/2$ (atmospheric), with $\D m^2 \gg \de m^2$. 
While propagating through ordinary matter, the coherent forward scattering of 
$\nu_e$ on electrons and nucleons differ from those of $\nu_\mu$ and $\nu_\tau$.
As a consequence, the oscillation probabilities are modified in a non trivial manner via the 
MikheyevÐ-SmirnovÐ-Wolfenstein (MSW) mechanism \cite{msw}.  Matter effects are specially 
significant in the Sun and other astrophysical objects and events, 
in particular, in core-collapse supernovae \cite{cvolpe}. 

Out of the six oscillation parameters,  $\de m^2$, $|\D m^2|$, and the mixing angles 
$\th_{12}, \th_{23}, \th_{13}$, are known at present \cite{concha}.  Thanks to the MSW 
effect involved in the flavor transformations of solar neutrinos we know that $\de m^2 > 0$ 
(i.e, $m_2 > m_1$).  The sign of $\D m^2$ depends on the ordering of the mass spectrum, 
positive for normal hierarchy $(m_3 > m_{1,2})$ and negative for inverted hierarchy 
$(m_3 < m_{1,2})$.  The data at hand neither allow to establish the ordering of the 
neutrino masses nor the value of the  phase $\de$ associated with a possible CP-violation 
in the leptonic sector. We also ignore if neutrinos are their own anti-particles. An intense 
research program is underway to address these and other important questions regarding 
neutrino physics \cite{caccianiga}.

Cosmic neutrinos offer opportunities both to understand the properties 
and behavior of neutrinos, as to probe the sources that produce them.  
Until now, two astrophysical objects have been observed with neutrinos, the Sun and 
the supernova SN1987A. High-energy extraterrestrial neutrinos have been observed by 
IceCube but their origin is still unclear \cite{icecube}.The detection of solar neutrinos 
not only confirmed that nuclear fusion reactions power the Sun, but also solved the 
solar neutrino puzzle, providing the only observed matter effect on the neutrino 
propagation to date \cite{maltoni}. On the other hand, observing supernova (SN)
neutrinos, both Galactic and relic, is in the agenda of the future large underground 
detectors \cite{dune}. As they stream out of a SN, in addition to the influence of the 
stellar matter (through the MSW effect), neutrinos are subject to the interaction 
with other neutrinos and antineutrinos \cite{notzold}.
The latter effect turns flavor evolution into a highly non-linear problem \cite{pantaleone} 
and gives rise to collective oscillations of the neutrino gas (see \cite{mirizzi} for recent 
reviews and extensive lists of references). This is a complex and rich phenomenon and, 
despite the significant progress made, a complete picture of it is still lacking. 
Limitations of the previous studies have been recently recognized and open issues 
pointed out \cite{chakraborty}. More realistic numerical studies with a full three-flavor 
mixing seem necessary to make further progress.

In a medium, the evolution equation of the neutrino flavor amplitudes is not
(in general) analytically solvable. In the case of ordinary matter, the way from collected 
data to sound determination of the values of the parameters requires massive numerical 
integrations of a linear, homogeneous system of ordinary differential equations (ODE) with 
coefficients depending on the distance neutrinos travel along a medium. However,  
there is a lack of discussion in the literature about the numerical methods used to this end.  
The efficiency of the procedure, i.e, the interplay between  numerical accuracy and computation 
time, seems to be given for granted.  Even though it is understandable, since it does not constitute an essential part of the research, is noteworthy the very few references existing in this regard in the specialized literature. To our knowledge, only two papers have addressed this issue, both of them within the realm of neutrino astrophysics \cite{duan, montecarlo}. Some articles 
in the same field have dealt with questions such as convergence and stability of the numerical solutions found by using ODE solvers \cite{numsol}, but no attention was paid to alternative algorithms.

The integration of the differential equations governing the matter neutrino oscillations
with classical general purpose methods of numerical integration (e.g. Runge\hs{.03cm}-Kutta schemes) does not preserve the norm of the solution at every integration step.  The value of the norm becomes rather a test of the quality of the integration itself. In contrast, the so-called 
geometric numerical integrators are not flawed by such an issue, since they are designed 
to preserve this property exactly (up to round-off error). They have been developed in recent 
years within the area of numerical analysis of differential equations \cite{arieh, hairer, casas}  
with the aim of preserving under discretization qualitative (very often geometric) properties 
the differential equation has. It is the main purpose of this work to argue in favor of their use 
as a convenient alternative to carry out neutrino oscillation numerical integrations.
Many geometric integrators exist but, to be concrete, we will focus on a particular class 
based on the Magnus expansion (ME) \cite{magnus, bcor_pr, bcor_ejp}. 
We explain shortly the way the algorithms are built up and give details that facilitate their coding. 
Other similar procedures could be applied with the same purpose without further complications.

ME provides an exponential representation for the solution of linear, homogeneous 
systems of differential equations.  Originally considered as a procedure to build  approximate 
analytical solutions of the time-evolution operator, in the sense of Perturbation Theory, its application as an efficient numerical integrator constitutes a relatively recent development \cite{arieh, bcor_pr}. Years ago, two of us introduced the use of the time-evolution operator to study matter neutrino oscillations and applied its representation in terms of the ME to incorporate non-adiabatic effects into the flavor transitions \cite{jajc, jcdo}. A distinctive feature of the ME is that the  {approximate solution provided by the procedure shares with the exact solution relevant geometric properties. More specifically, for the case under consideration, the relevant property} corresponds to the unitary character of the time-evolution operator and, consequently, the norm of the solution vector. In a numerical calculation, this is verified at every single integration step 
and, therefore, probability is conserved by the final solution obtained through the composition 
of the successive steps. As already mentioned, this characteristic is not shared by classical  
numerical integrators like, for instance, those of the Runge-Kutta class.  Another, more practical, advantageof a ME based method is the use of larger values of the integration step which conveys shorter CPU time to the same solution accuracy.

The paper is organized as follows. In Section \ref{sec:equation} we write down the neutrino flavor evolution 
equation to be integrated. The basics of the ME are given in Section \ref{sec:basics}, and their implementation 
in the problem at hand is carried out in Section \ref{sec:numerics}.  Section \ref{sec:results} contains the 
outputs of the integration for three neutrino oscillations in the Sun and a supernova, modeled by simple 
density profiles. A comparison in terms of the computing efficiency is also provided for an algorithm based
on ME and a standard integration routine. The discussion of the results is in Section \ref{sec:discus}.

\section{The flavor evolution equation} \label{sec:equation}

Except for some anomalous results from short baseline measurements  
which require further clarification \cite{anomal}, existing data can be 
interpreted with the simplest extension of the standard model to incorporate
nonzero masses of mixed active neutrinos.  Namely, within a framework 
where the three known flavor states 
$|{\nu}_{\al}\ran (\al = e, \mu, \tau)$ are linear combinations 
of the states $|{\nu}_i \ran$ with masses $m_i (i= 1,2,3)$: 
\be
\label{flav&massrel}
|{\nu}_{\al}\ran= \sum_i U^*_{\al i} |{\nu}_i \ran.
\ee
The coefficients $U_{\al i}$ are elements of the unitary mixing matrix  $U$ that
appears in the charged current, the so called Pontecorvo--Maki--Nakagawa--Sakata 
(PMNS) matrix. For Dirac neutrinos  this matrix is customarily expressed as
\be
\label{eq:Ufactorization}
U = \Oyz\G\Oxz\G^\dagger\Oxy\,,
\ee
\ni
where the orthogonal matrices ${\cal O}_{ij}$ represent rotations by angles
$\th_{ij}\!\in\![0, \pi/2]$ in the respective planes, while $\G = diag \hs{.03cm} (1, 1, e^{i\de})$,
with $\de\!\in\![0, 2\pi]$. If neutrinos are Majorana particles $U$ has to be multiplied by the right by 
another diagonal matrix $\G_{\!\rm\chic{M}} =  diag  \hs{.03cm} (1, e^{i\de_1}\!, e^{i\de_2})$. 
The two additional physical phases do not play a role in neutrino oscillations and are therefore omitted
in the analysis of the phenomenon \cite{bilenky}.

Let us  consider a neutrino $\nu_\al$  produced within a medium at time $t_0$.
The state of the system $|\psi (t)\ran$ at time $t \geq t_0$ can be expressed as
$|\psi (t) \ran = \sum_\b \psi_\b(t)|\nu_\b\ran$, with $\psi_\b(t_0) = \de_{\al \b}$.
The probability to have a state of flavor $\b$ at a point $ r = t\,(\hbar=c=1)$ is
\be
\label{oscprob1}
P_{\al \b} (r) = |\psi_\b (r)|^2 .
\ee

Once the neutrinos leave the medium the amplitudes evolve according to the equation 
that governs vacuum oscillations, whose solution is simpler when written in the basis of 
the mass eigenstates. According to Eq. \eqref{flav&massrel}, denoting by
$
{\mc A}_j = \phi_j(r_{\!\star})
$ 
the probability amplitude of having a $\nu_j$ at the edge of the medium, 
for $r \geq r_\star$ we can write
\be
\label{flavoramp}
\psi_\b (r)=\sum_{j=1}^3 U_{\b j} {\mc A}_j \exp(-i E_j L)\,,
\ee
where 
$E_j = \sqrt{|\mb p|^2 + m_j^2}$ and $L = r - r_\star$ is the distance travelled by the
neutrinos in vacuum.
Substituting \eqref{flavoramp} into Eq. \eqref{oscprob1} we obtain
\bea
\label{oscprob2}
P_{\al \b} && \hs{- 0.3 cm} =  \sum_j |U_{\b j}|^2 |{\mc A}_j|^2 \non \\ 
&& + \,2\sum_{i>j} \mrm{Re}\left[
U_{\b i} U^*_{\b j} {\mc A}_i {\mc A}^*_j\exp\left(-i \D_{ij}L\right)\right].
\eea
The quantity $\D_{i j}= \D m^2_{i j}/2E$, with $E = |\mb p|$, is the oscillation 
wave number associated with the squared mass difference $\D m^2_{ij}=m_i^2-m_j^2$.

As a result, the problem of calculating $P_{\al \b}$ reduces to determine the quantities 
${\mc A}_j$, i.e., find the mass eigenstate amplitudes $\phi_j(r)$ within the medium, subject to the 
initial condition 
$
\phi_j(r_0) =  U^*_{\al j}.
$
For relativistic neutrinos propagating in normal matter, after substracting a  
global phase, the evolution equation for these amplitudes
has the form
\be
\label{Scheq}
i \frac{d}{dr}
\Phi(r)= \big[ H_0 +  v(r)\hsp U^\dag\hspace{.01cm} V
 U \big]
\hs{-0.02 cm} \Phi(r)\,,
\ee
with 
$\Phi^{\mtt T}(r) = \big( \phi_1(r), \phi_2(r), \phi_3 (r)\big)$,
$U$ the mixing matrix, and $V = {\rm diag}\,(1,0,0)$.
The first term between brackets is the Hamiltonian matrix that governs 
the flavor evolution in vacuum $H_0 = {\rm diag}\,(0,\D_{21},\D_{31})$,
while the second term accounts for the matter effects due to the coherent 
interaction of  the neutrinos with the background particles. The quantity 
\be
v (r) = \sqrt{2}\hsp G_{\hs{-.03cm}F} n_e(r)
\ee
denotes the difference between the effective potential energies of 
the $\nu_e$ and the $\nu_{\mu,\tau}$. Here, $G_{\hs{-.03cm}F}$ 
is the Fermi constant and $n_e(r)$ is the number density of electrons 
along the neutrino path. 

If neutrinos propagate only in vacuum, then  $v(r) = 0$ along their entire way 
to the detector and $L$ corresponds to the distance from the 
production point at $r_0$. Replacing ${\mc A}_j$ by $U^*_{\al j}$ in 
Eq. \eqref{oscprob2} we recover the usual formulas for the oscillation 
probabilities in vacuum.

Since $\Oxy$ and $\G$ commute, the PMNS matrix can be written as
$
U = \Oyz\G\mc O\hsp\G^\dag,
$
where $\mc O =  \Oxz \Oxy$. From this, taking into account that the 
commutators $[V,{\Oyz} ], [V,{\G}]$,
and $[H_0,{\G}]$ vanish, we obtain
\be
\label{Scheq2}
i \frac{d}{d r}{\Psi}(r) = H(r) {\Psi}(r)\,,
\ee
with ${\Psi}(r) = {\G}^\dag \Phi(r)$ and 
\be  \label{eq:H1}
H (r) =  H_{\rm 0} + v(r) W\,.
\ee
The matrix $W$ is given by 
$
W = {\mc O}^{\mtt T}V \mc O, 
$
where ${\mc O}^{\mtt T}$ is the transpose of the orthogonal matrix $\mc O$. 
Explicitly,
\be 
\label{eq:matrixW}
W = \left(\!
\begin{array}{ccc}
c_{\xz}^2c_{\xy}^2 & c_{\xy}s_{\xy}c_{\xz}^2 & c_{\xy}c_{\xz}s_{\xz} \\
c_{\xy}s_{\xy}c_{\xz}^2 & s_{\xy}^2c_{\xz}^2 & s_{\xy}c_{\xz}s_{\xz} \\
c_{\xy}s_{\xz}c_{\xz} & s_{\xy}c_{\xz}s_{\xz} & s_{\xz}^2
\end{array}
\!\right)\!,
\ee
with $s_{ij}=\sin \theta_{ij}$ and $c_{ij}=\cos \theta_{ij}$. 
Note that $H(r)$ is a real and symmetric matrix that can be diagonalized by an 
$r$-dependent orthogonal transformation. It does not contain  $\th_{23}$ nor $\de$
and then its eigenvalues, which are identical to those of the Hamiltonian in Eq. \eqref{Scheq},
do not depend on these  two parameters.

Typically, the oscillatory interference terms in \eqref{oscprob2} 
average to zero for neutrinos  traveling a long distance to the Earth.
Under such circumstances, the average probability of finding a $\nu_\b$ at 
a detector in the Earth is given by the incoherent superposition
\be
\label{avosprob}
\lan P_{\al \b}\ran = \sum_j  |U_{\b j}|^2 {\mc P}_j\,,
\ee
where ${\mc P}_j \equiv |{\mc A}_j|^2 $ denotes the  
probability to have a mass eigenstate at the surface of the star.
Neutrinos that travel a certain distance through the Earth before reaching 
the detector will experience oscillations in the terrestrial matter. 
As a consequence, $U_{\b i}$ in $\lan P_{\al \b}\ran$ has to be replaced
by the amplitude for a $\nu_i$ emerging from the Earth be in 
a flavor state $\nu_{\b}$. We do not consider this effect here but, 
whenever needed, it can be easily incorporated in our approach.

If the initial  state corresponds to an electron neutrino, then
\be \label{eq:IC}
\Psi(r_0)=
\left(\!
\begin{array}{c}
       c_{\xy} \, c_{\xz} \\
       s_{\xy} \, c_{\xz} \\
       s_{\xz}
\end{array}
\!\right)
\ee
and, according to Eq. \eqref{avosprob}, the average survival probability reads
\be
\label{eq:avsurvprob}
\lan P_{ee} \ran = 
c_{\xy}^2 c_{\xz}^2 {\mc P}_1 + s_{\xy}^2 c_{\xz}^2 {\mc P}_2 + s_{\xz}^2 {\mc P}_3\,,
\ee
where 
\be
\label{eq:probconserv}
\sum_{j = 1}^3{\mc P}_j = 1
\ee
by conservation of probability. For solar neutrinos the flux of 
electron neutrinos at the detection point $F_{\nu_e}$ is obtained multiplying
the original flux $F_{\nu_e}^{(0)}$ by the expression of $\lan P_{ee}\ran$
as a function of energy. 

On the other hand, neutrinos (and  antineutrinos) 
of all species are produced during a core-collapse SN.
The flux of $\nu_e$ arriving at Earth may be written as
\cite{dighe&smirnov}
\be
\label{probSN}
F_{\nu_e} = F^0_{\nu_e} \lan P_{ee}\ran+ F^0_{\nu_x}(1 - \lan P_{ee}\ran)\,,
\ee
where $\nu_x$ stands for either $\nu_{\mu}$ of ${\nu_\tau}$. In the deepest 
regions the neutrino density is now so high that neutrino-neutrino interactions 
are dominant and produce novel collective effects \cite{mirizzi}.  The net result
is the modification of the primary fluxes $F^0_{\nu_\al}$ because of spectral swaps 
in certain energy intervals. Usually, this happens within the first 
few 100 km, much before neutrinos experience the MSW conversions we consider
here, which can be incorporated separately.

In the rest of the article, we will focus on the determination of 
{the vector $\Psi(r)$ or equivalently}
the quantities
${\mc P}_j$ calculated numerically by means of the Magnus procedure. When substituted into 
Eq. \eqref{eq:avsurvprob} they render us $\lan P_{ee}\ran$ as a function of
the energy.  The corresponding curves for the Sun and a SN are plotted
in Figure \ref{Fig-P}. They were obtained by means of the method denoted 
as M4 in Section \ref{sec:numerics} and using a set of the oscillation parameters
consistent with the experimental results.

\section{Magnus expansion: Basics} \label{sec:basics}

For the sake of completeness we present here the basics of ME \cite{bcor_ejp}. 
The explanations are oriented toward its use as a numerical integrator.

Given the matrix differential equation
\begin{equation}   
\label{eq:evolution}
   \frac {d }{dt}Y(t) = A(t) Y(t), \qquad A \in \mathbb{C}^{\hs{.003cm}\mtt n \times \mtt n},\quad Y \in 
   \mathbb{C}^{\hs{.005cm}\mtt n}, 
\end{equation}
with the initial condition $Y(t_0)=Y_0$, the Magnus  {approach} consists in looking for an exponential 
representation of the solution 
\begin{equation} 
\label{eq:exprep}
Y(t) = \exp\hs{-.03cm}\big\{\Omega(t,t_0)\!\big\}Y_0\,,
\end{equation}
with
\begin{equation} 
\Omega (t,t_0) \in \mathbb{C}^{\hs{.003cm}\mtt n \times \mtt n}, \qquad \Omega(t_0,t_0)=0\,.
\end{equation}

When the matrix elements of $A$ are $t$--independent, then $\Omega (t, t_0) = (t-t_0)A$. 
In general, $\Omega(t,t_0)$ is given as a series expansion
\begin{equation} 
\label{eq:series}
\Omega(t,t_0)=\sum_{k = 1}^\infty \Omega_k(t,t_0), \qquad \Omega_k(t_0,t_0)=0\,,
\end{equation}
where the term $\Omega_k$ is built up from  {multiple} integrals of 
nested commutators of $k$ matrices $A$ evaluated at $k$ different times \cite{magnus}. 
The complexity of the individual terms  {largely} increases with the index $k$. 
In particular, the first three ones read
\bea
\label{O1}
  \Omega_1(t,t_0) &=&\int_{t_0}^t {\rm d}t_1 A _1\,, \\
\label{O2}
  \Omega_2(t,t_0)&=&\frac{1}{2}\int_{t_0}^t {\rm d}t_1 \int_{t_0}^{t_1} {\rm d}t_2
  [A _1,A _2]\,,\\
\label{O3}
\Omega_3(t,t_0)&=&\frac{1}{6}\int_{t_0}^t {\rm d}t_1 \int_{t_0}^{t_1}{\rm d}t_2
  \int_{t_0}^{t_2} {\rm d}t_3 \non\\ 
  \phantom{\Omega_3(t,t_0)}&& \big\{[A _1,[A _2,A _3]]+[[A _1,A _2],A _3]\big\}\,, \label{eq:Om3}
\eea
where $A_i \equiv A(t_i)$ and the square brackets stand for the commutator:
$[A,B]\equiv AB-BA$, for $A,B$, matrices of appropriate dimensions.

For the purposes of using ME as a numerical integrator that furnishes a solution $Y(t_f)$ 
starting from $Y_0$, the question focuses on how to handle efficiently
a single integration step. Namely,  one considers certain  grid points
$t_0 < t_1 < \cdots < t_N = t_f$ of the time interval $[t_0, t_f]$ with associated time increments $h_n = t_{n+1} - t_n$ for
$0 \le n \le N-1$ and then determines
\begin{equation}
\label{eq:step}
Y(t_{n+1})=\exp\hs{- 0.03 cm}\big\{\Omega (t_n+ h_n ,t_n)\!\big\} Y(t_n)\, .
\end{equation}
After that, iteration yields the solution in $N$ steps
\begin{equation}
\label{eq:iteration}
Y(t_f )=\prod_{n=0}^{N -1} \exp \left\{ \Omega (t_n; h_n)\right\} Y_0\,,
\end{equation}
with $\Omega (t_n; h_n) \equiv \Omega (t_n+ h_n, t_n)$.
When the matrix $A$ is anti-Hermitian then the matrix $\prod_{n=0}^{N -1} \exp \left\{ \Omega (t_n; h_n)\right\}$
constitutes a unitary approximation to the time-evolution operator between $t_0$ and $t_f$, since it is the product of $N$ 
unitary matrices.

The procedure has to face three problems of different nature:
\bei

\item
Truncate the series (\ref{eq:series}) for $\Omega (t_n; h)$ at an appropriate index,
\be
\label{eq:truncseries}
\Omega^{(p)} (t_n; h_n) = \sum_{k = 1}^p \Omega_k (t_n; h_n)\,.
\ee

\item
\vs{-.1cm}
Generate the quadratures to  {approximate} the multivariate integrals
in the truncated series.

\item
Compute the exponential of  {the resulting} square matrix. 

\eni

\ni
The last point will be treated later on and an explicit algebraic form given for the 
three neutrino case. The other two are solved at once and depend on the order 
of approximation of the method. This is  defined as the order $r$ of the Taylor 
expansion that gives  $Y(t_n + h_n )$, up to $\emph{O}\,(h_n^{r+1})$, out of the 
exact value $Y(t_n)$.  One important point is that the  {order of the quadratures required  
in the} numerical approximation of $\Omega_k$  in the truncated series (\ref{eq:truncseries}) 
is not greater than the order of approximation $r$. This fact is 
exploited so as to minimize the number of operations in the algorithm. Moreover, 
since ME is time symmetric, for achieving an integration method 
of order $2 r$ (with $r > 1$) only terms up to $p=2r - 2$ are required in the series expansion (\ref{eq:series}).
Book-keeping technicalities may be found in the literature \cite{arieh, bcor_pr}.

\section{Numerical integration} 
\label{sec:numerics}

The differential system that rules the evolution of the three-neutrino 
amplitudes in matter (Eq.~(\ref{Scheq2})) is of the type in (\ref{eq:evolution}) with $\mtt n = 3$,  $t = r$, and 
the identifications
\begin{equation}
A\equiv -iH , \qquad Y\equiv \Psi\,.
\end{equation}
Now, we apply the approach outlined in the previous section to solve numerically
such equation with the initial condition (\ref{eq:IC}). Since each term in 
the series expansion of ${\Omega(t, t_0)}$ is anti-Hermitian, ME provides 
a representation of $\exp\hs{-.03cm}\big\{\Omega(t,t_0)\!\big\}$ in \eqref{eq:exprep}
that is unitary  {by construction, even if the series is truncated}.
This property is preserved by the factorized expression given in Eq. \eqref{eq:iteration} and 
by the algorithm used to compute each factor. 
As a consequence, condition \eqref{eq:probconserv} is fulfilled automatically
by the ${\mc P}_j$ calculated using  the procedure. 

We assume normal hierarchy and adopt the best-fit values of the 3-neutrino oscillation 
parameters, derived from a global fit of the  current neutrino data \cite{concha}:
$\D m_{21}^2 = 7.54 \times 10^{-5} eV^2, \D m_{31}^2 = 2.47 \times 10^{-3} eV^2$,
$\sin^2\th_{12} = 0.308, \sin^2\th_{23} = 0.437$, 
and $\sin^2\th_{13} = 0.0234$. In addition, 
the distance is expressed in units of the solar radius $R_\odot = 6.96 \times 10^5$ km. 
Thus, written in terms of the  dimensionless variable $\xi \equiv r/R_\odot$, Eq. \eqref{Scheq2} becomes
\begin{equation} 
\label{eq:Hme}
i\frac{\mathrm{d} \Psi}{\mathrm{d}\xi} = \big[H_0+v(\xi)W\big]\Psi,
\end{equation}
with $W$ the matrix given in \eqref{eq:matrixW} and
 \begin{equation}
 \label{eq:H0Y}
	H_0=\frac{a}{\mc E}\left(
	\begin{array}{ccc}
	0 & 0 & 0 \\ 
	0 & b & 0 \\ 
	0 & 0 & 1
	\end{array}  
	\right) .
\end{equation}
Here ${\mc E}$ is the numerical value of the neutrino energy in MeV and 
$a= 4.35196 \times 10^6$ and $b= 0.030554$ are dimensionless
parameters. 

The conclusions about the numerical efficiency are {essentially} independent
on the initial integration point. To be concrete, for the computations we use $\xi_0 = 0.1$ 
in the case of the Sun and $\xi_0=0.02$ in the case of a SN. We also take the supernova's radius  equal to $20\hs{.03cm}R_\odot$. Accordingly, the integration yields 
$\Psi(1)$ and $\Psi(20)$ for the Sun and a SN, respectively.
The integration procedure requires a particular shape of the function
$v(\xi)$ to be provided. In order to keep as clear as possible the technical explanations 
we consider two simple profiles of the electron number density existing in the 
literature. For the Sun, we model the density by an exponential profile 
$v(\xi)=\gamma \exp(-\eta\hs{.03cm}\xi)$, with $\gamma=6.5956 \times 10^4$ and $\eta =10.54$
\cite{bahcall}. In the SN case, we adopt the power law used in \cite{sunova}, namely, 
$v(\xi)=\gamma/\xi^3$, with $\gamma=52.934$. Let us remark that $v(\xi)$ does not need to be 
analytically defined, nor to be a continuous function, nor to have continuous derivatives. 
This applies, for instance, to neutrinos that traverse the Earth, which radial matter density 
is represented by a discontinuous piecewise function according to the
Preliminary Reference Earth Model (PREM) \cite{prem}.

Next, we describe the main features of the numerical integration of Eq.~(\ref{Scheq2}) by means of  two particular schemes based on the ME of orders 2 and 4.
Starting the integration at $\xi_0$, with the initial value $\Psi(\xi_0)$ we build up the approximate solution 
at $\Psi(\xi_1 = \xi_0+h_1)$. At a generic point $\xi_n$, the solution $\Psi(\xi_n)$
determines $\Psi(\xi_n + h_n)$ as 
\be
\label{eq:eOp}
\Psi(\xi_{n+1})  =  \exp\left\{ \Omega^{[r]}(\xi_n;h_n) \right\} \Psi(\xi_n).
\ee
 {Here $\Omega^{[r]}$ denotes an approximation to the truncated Magnus series (\ref{eq:truncseries}) of order $r$ in $h_n$}, 
with $r = 2, 4$. The iteration stops  after $N$ iterations at $\xi=1,20$.

\subsection{Order 2 formula: M2}

The method of order $r = 2$ is particularly simple for two reasons. 
First, just the term \eqref{O1}, namely,
\be
\Omega_1(\xi_n; h_n) = - i \int_{\xi_n}^{\xi_n + h_n}\hsp {\rm d}t\hsp H(t),
\ee
 {has to be taken into account} in the series (\ref{eq:truncseries}).
Second, the quadrature for this  {integral} requires only one evaluation point 
to obtain a second order  {approximation}. 
Thus, for the exponent in \eqref{eq:eOp} we get
\begin{equation} 
\label{eq:vt12}
\Omega^{[2]}(\xi_n;h_n)= -i H(\hsp\bar \xi \,)\hsp h_n=-i\big(H_0+ \bar v \hsp W\big)h_n\,, 
\end{equation}
with $\bar \xi \equiv \xi_n + {h_n}/{2}$. The quantity $\bar v \equiv v(\bar \xi \,)$ must be 
re-evaluated at every step. This scheme is also known as the exponential midpoint rule 
and we will refer to it as M2.

\subsection{Order 4 formula: M4}

 {With $p=2$ in the truncated series (\ref{eq:truncseries}) one ends up with a method of 
order $r = 4$ if the corresponding integrals
 \eqref{O1} and \eqref{O2} are approximated by a 2-point Gauss--Legendre quadrature rule \cite{arieh, bcor_pr}. Specifically,
given the quadrature points}
\be
\xi_{\pm}= \xi_n + \left(1 \pm \frac{1}{\sqrt{3}}\right)\!\frac{h_n}{2}\,,
\ee
with $\xi_{-} < \xi_{+}$ and defining the quantities
\begin{equation}
\label{AiG4}
     H_{\pm} =  H(\xi_{\pm})\,,
\end{equation}
the one-step integration exponent
of order $4$ can be cast into the form
\begin{equation} 
\label{eq:O4}
\Omega^{[4]}(\xi_n;h_n) = - i \big(H_{+} + H_{-}\big) \frac{h_n}{2} + 
          \frac{\sqrt{3}}{12}[H_{-}, H_{+}]h_n^2\,,
\end{equation} 
where, as before, the square brackets stand for the matrix commutator. 
Alternatively, the Simpson quadrature rule also yields an equivalent fourth-order approximation
\cite{bcor_pr}.

Working out equation (\ref{eq:O4}) for the matrix $H$ given in (\ref{eq:H1}), we arrive at
\begin{eqnarray}
\label{eq:HO4}
\Omega^{[4]}(\xi_n;h_n)= \!&-&\!i\Big(H_0 + \frac{1}{2}(v_+ + v_-)W\Big)h_n \nonumber \\
&+&	\! \frac{\sqrt 3}{12} (v_{+} - v_{-}) \,  [H_0, W] \, h_n^2\,,
\end{eqnarray}
where $v_\pm \equiv v(\xi_\pm)$, are the only quantities to be re--evaluated 
at every integration step. The matrix $[H_0,W]$ is built up only once, namely 
at the beginning of the integration process. 
This method will be referred to as M4. 

\subsection{Variable step size implementation}
\label{vssi}

The easiest way to implement the previous methods is by considering a constant step size, i.e., by taking $h=(\xi_f - \xi_0)/N$ and then
setting $\xi_n = \xi_0 + n h$. This implementation tends to be inefficient, however, since the solution $\Psi(\xi)$ may experience rapid changes
along the evolution on some intervals and evolve slowly on some others. Thus, it is better to adjust $h_{n}$ accordingly as the integration proceeds.
There are several techniques for doing this automatically in such a way that the local error is below a prescribed tolerance \texttt{tol}, one of the most
common being the local extrapolation procedure \cite{hairer2,sanz-serna}. In our setting this approach can be summarized as follows.

We produce two numerical solutions at $\xi_{n+1}$  according with M2 and M4 above,
\begin{equation}
  \hat{\Psi}_{n+1} = e^{\Omega^{[2]}(\xi_n,h_n)} \Psi_n,  \qquad    \Psi_{n+1} = e^{\Omega^{[4]}(\xi_n,h_n)} \Psi_n,
\end{equation}
respectively. Then the quantity 
\begin{equation}  \label{estim}
  E_r = \| \hat{\Psi}_{n+1} -  \Psi_{n+1} \|
\end{equation}  
   can be used to estimate the local error corresponding to M2. If $E_r$ computed
at $\xi_{n+1}$ is below \texttt{tol}, then the step from $\xi_n$ to $\xi_{n+1}$ is accepted and then we proceed to compute the approximation
to the solution at $\xi_{n+2}$. If $E_r > \mbox{\texttt{tol}}$, then the approximation at $\xi_{n+1}$ is rejected and a smaller step is chosen to
compute a new approximation at $\xi_{n+1}$. In either case, the value of the new step is given by \cite{sanz-serna}
\begin{equation}  \label{new.step}
  h_{\mathrm{new}} = s \, h_c \left( \frac{\mbox{\texttt{tol}}}{E_r} \right)^{1/3},
\end{equation}
where $h_c$ denotes the current value of the step size and $s$ is a ``safety factor" chosen to decrease the probability of a rejection at the next step.
For our problem a good choice is $s = 0.8$. Notice that once the step $\xi_n \rightarrow \xi_{n+1}$ has been completed, we have two numerical 
approximations at $x_{n+1}$:  $ \hat{\Psi}_{n+1}$ and $\Psi_{n+1}$, so that it is possible to choose one or the other as the approximation to
$\Psi(\xi_{n+1})$. In local extrapolation, one advances with the higher-order result ${\Psi}_{n+1}$ (hence the name).

Since the most time-consuming part of the Magnus methods corresponds to the computation of the exponential, evaluating directly $E_r$ as
(\ref{estim}) may increase considerably the total computational cost of the algorithm. To avoid that we can express
\begin{equation}
   \hat{\Psi}_{n+1} -  {\Psi}_{n+1}   = \left( e^{\Omega^{[2]}} - e^{\Omega^{[4]}} \right) \Psi_n = ( e^Z - I)  {\Psi}_{n+1},
\end{equation}
where
\begin{equation}
  Z = \log( e^{\Omega^{[2]}} e^{-\Omega^{[4]}} ) = \Omega^{[2]} - \Omega^{[4]} - \frac{1}{2} [\Omega^{[2]}, \Omega^{[4]}] + \cdots.
\end{equation}
Working out this expression and expanding $e^Z$, we arrive at
\begin{equation}  \label{estim.2}
   E_r = \|(h_n^2 S_1 + h_n^3 S_2 + \frac{1}{2} h_n^4 S_1^2) {\Psi}_{n+1}\| + \mathcal{O}(h_n^5),
\end{equation}
where
\begin{eqnarray}
 S_1 & = &  -\frac{\sqrt{3}}{12} (v_{+} - v_{-}) \, [ H_0, W] \nonumber \\
 S_2 & = & i \frac{\sqrt{3}}{24}  (v_{+} - v_{-}) \Big( [H_0,[H_0,W]]  \\
   & &  + \frac{1}{2} (v_{+} + v_{-}) \, [W,[H_0,W]] \Big). \nonumber
\end{eqnarray}
Notice that estimating $E_r$ according to (\ref{estim.2}) only requires additional combinations of $v_\pm$ at each step and the evaluation of
the nested commutators $[H_0,[H_0,W]]$ and $[W,[H_0,W]]$ at the beginning of the integration.
In practice, we will scale the $i$-th component of  $\hat{\Psi}_{n+1} -  {\Psi}_{n+1}$ by a factor $d_i = |({\Psi}_{n+1})_i |$ to work with relative
errors. As for the initial step size, a possible choice is just $h_0 = \mbox{\texttt{tol}}/2$.

More sophisticated schemes exist, of course, for step-size control that are employed in commercial  computer packages. The previous method,
although simple, is quite efficient and well adapted for the problem at hand.

As with respect to the computation of the matrix exponential, 
the Appendix below provides an analytic formula for that.
It has been written down so as to cope with eventual quasi--degeneracy of eigenvalues (causing extra rounding-off errors) and to allow minimal amount of arithmetic. 
Its direct coding is straightforward and saves computing time with respect to general purpose routines developed for generic dimension matrices.

\section{Results} \label{sec:results}

All the result we will show correspond to outputs of double precision computations carried out in Fortran with no optimization during the compilation. They have been replicated in Matlab albeit investing longer CPU times.

To test the performance of the M4 method we have carried out the very same computations using the well known ODE solver Dopri5(4) designed by Dormand and Prince \cite{dormand}. It is an  adaptive step-size explicit 
Runge\hs{.03cm}-Kutta method of order five that has embedded a  fourth-order approximation used to estimate the error and applies local extrapolation. The Matlab procedure \texttt{ode45}, in particular, is based on this
integrator.

Here we have used the standard Fortran implementation of Dopri5(4) called DOPRI5 available at \cite{dopri5}. 
Given a particular level of accuracy, the focus will be on the relatively shorter CPU time needed by M4 in comparison with DOPRI5, rather than on the absolute measures of CPU time. It might be, however, of interest to point out that computations were carried out on an Intel Core Duo E8400 
processor running at 3 GHz. 

\begin{figure}
\begin{center}
\includegraphics[width=0.45\textwidth ,height=!,angle=0]{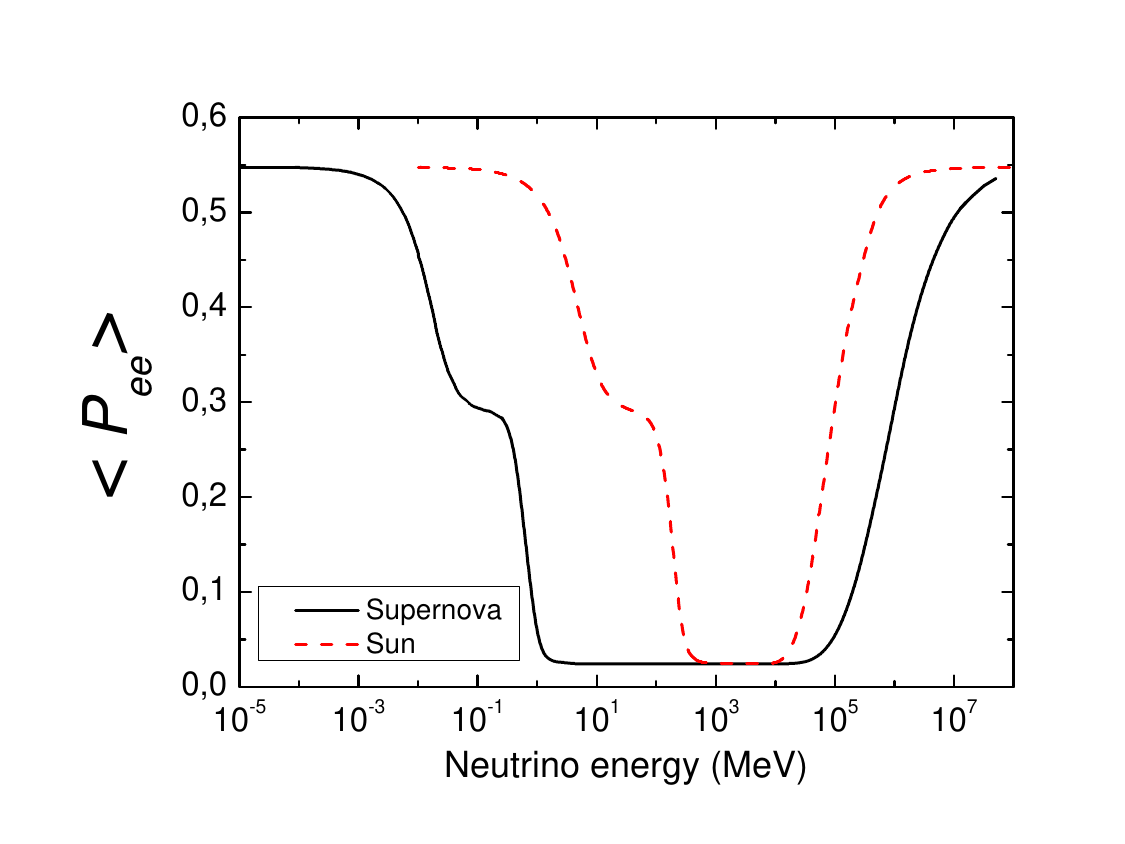}
\end{center}
 \caption{Survival probability for electron neutrinos, Eq.~(\ref{eq:avsurvprob}), at the edge of the Sun (dashed line) and a supernova (solid line) as a function of the neutrino energy $E$. The curves have been obtained with the Magnus solver M4.}
 \label{Fig-P}
\begin{center}
\includegraphics[width=0.45\textwidth ,height=!,angle=0]{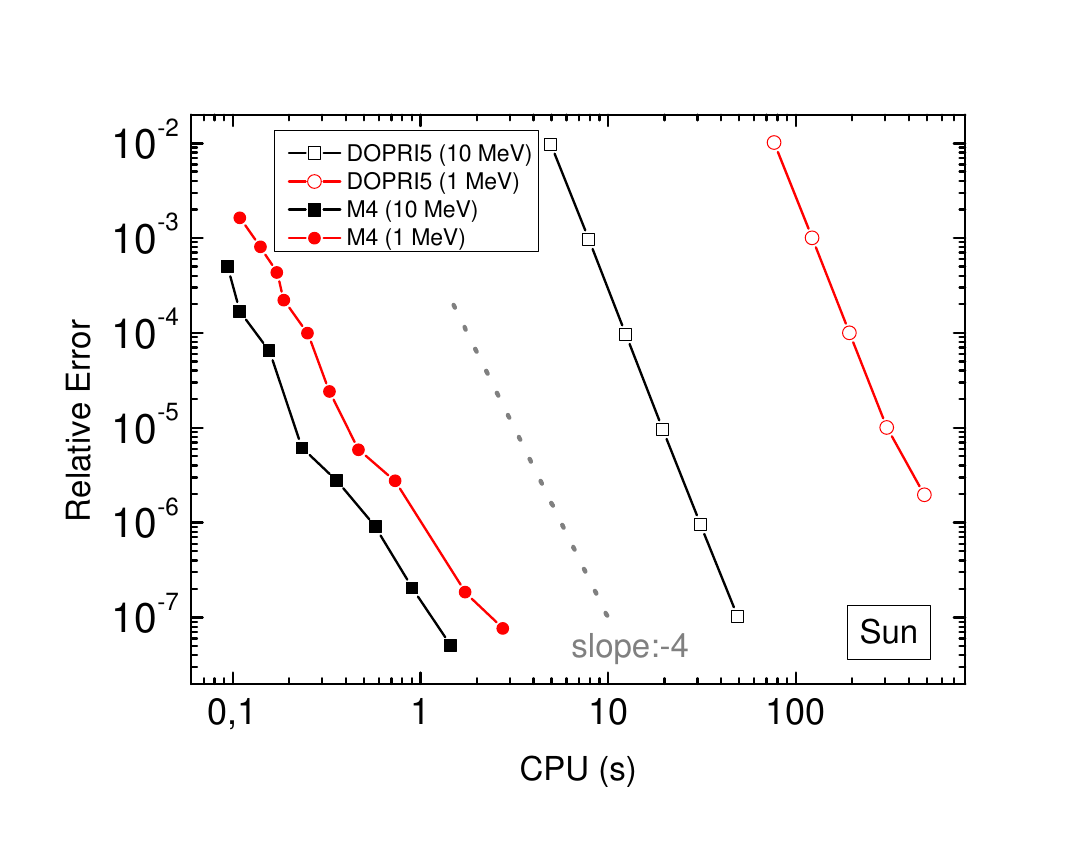}
\end{center}
 \caption{ Relative error of the solution of Eq. (\ref{eq:Hme}) as a function of the CPU cost for solar neutrinos with energies $E =1$ MeV (circles) and $E =10$ MeV (squares). Numerical integrations were carried out with the solvers 
M4 with variable stepsize (solid) and DOPRI5 (open).}
 \label{Fig-Suncpu}

\begin{center}
\includegraphics[width=0.45\textwidth ,height=!,angle=0]{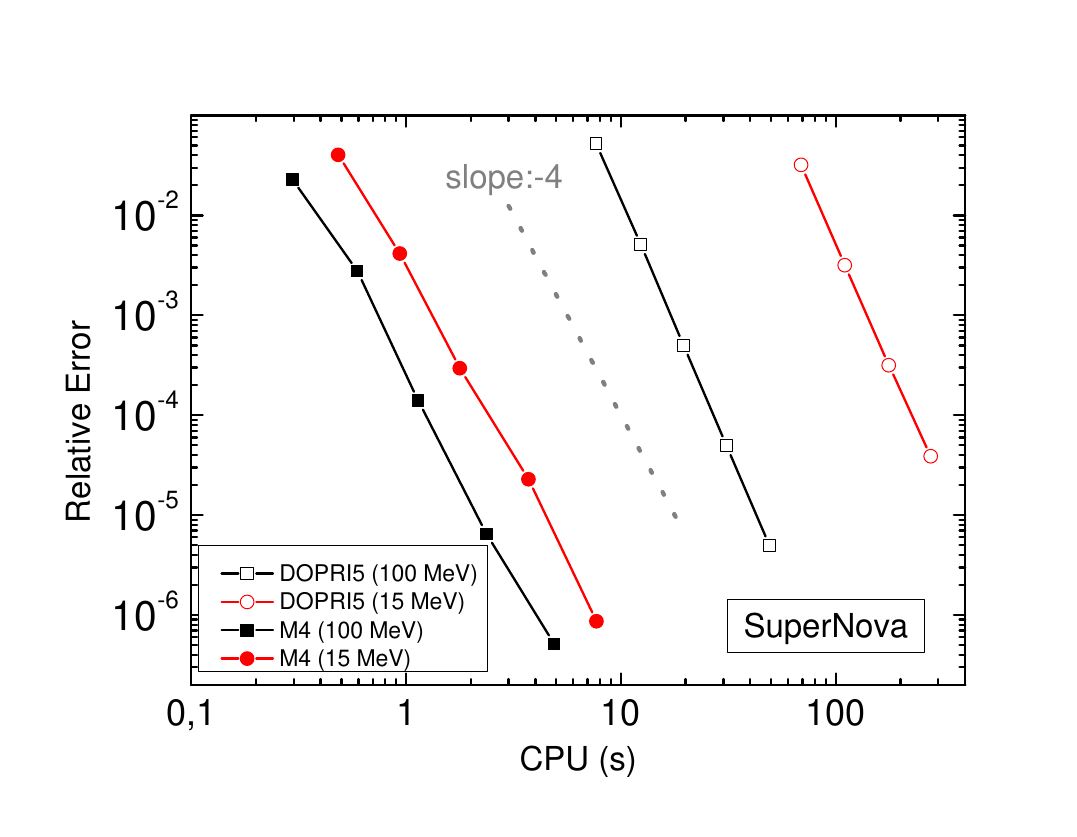}
\end{center}
 \caption{ Relative error of the solution of Eq. (\ref{eq:Hme}) as a function of the CPU cost for supernova neutrinos with energies $E=15$ MeV (circles) and $E=100$ MeV (squares). Numerical integrations were carried out with the solvers M4 (solid) and DOPRI5 (open).}
 \label{Fig-SNcpu}
\end{figure}

In Figure \ref{Fig-P} we plot the average electron-neutrino survival probability (Eq.(\ref{eq:avsurvprob})) 
at the edge of Sun (dashed line) and a SN (solid line) as a function of the neutrino energy.
These curves reproduce results of the same kind available in a number of references.
As a test, we mention that the theoretical asymptotic value ${\langle P}_{ee} \rangle = 0.547829$ 
is correctly reproduced, both in the high energy and the low energy limits.

Figures \ref{Fig-Suncpu} and \ref{Fig-SNcpu} contain the main output of the present work. They show the relationship between the accuracy obtained for a numerical solution and the  {computational cost required as measured by the} CPU time invested to achieve it. Every CPU time corresponds to a particular value of the tolerance $\texttt{tol}$. The smaller $\texttt{tol}$, the larger CPU time. 
These curves allow one to compare the relative efficiency of the two ODE solvers tested: M4, and DOPRI5.
Two different representative values of the neutrino energy have been considered,  as specified in each diagram.
Given a value for the relative error (to be defined in next paragraph) of the numerical solution of \eqref{eq:Hme} then the corresponding abscissa of every curve determines the CPU time that was invested in the computation from $\xi=0.1$ up to $\xi=1$ (Sun) and from $\xi=0.02$ up to $\xi=20$ (SN).  {Certainly,} the CPU time is not an absolute measure of efficiency since it depends on the type of processor, the programmer expertise, and the compiler. Nevertheless, 
the location of the curves in the plots is a reasonable relative measure of efficiency of each method, provided  the runs are carried out on the same computer and compiler. This type of plots allow us to visualize the balance between the accuracy required in the solution and the affordable numerical effort. 

In order to build up the relative error we employed an ``exact" numerical solution 
$\Psi^{ref}(\xi)$ obtained by means of the command \texttt{NDSolve} of \emph{Mathematica}
with stringent requirements to guarantee high accuracy. Specifically, 
the relative error is determined as the norm of the vector with components given by the quantities
$
(\psi_j(\xi) - \psi_j^{ref}(\xi))/ \psi_j^{ref}(\xi) (j=1,2,3),
$
evaluated at  $\xi=1$ (Sun) or $\xi = 20$ (SN). Notice that massive numerical computations cannot afford the huge amount of CPU time required  by the procedure to obtain $\Psi^{ref}(\xi)$. Hence, the need for efficient ODE solvers. 

When one plots the error obtained with an integrator of order $r$ as a function of the computational cost in a double logarithmic scale, in the limit $h_n \rightarrow 0$ one obtains 
a straight line with slope $-\hs{.03cm}r$, so that usually the order of the method can be deduced
at once from the diagram.
For this reason, we have included the straight line with  slope $-4$, as visual reference.  

Given that $\mathcal{E}$ appears in the denominator of (\ref{eq:H0Y}) the instantaneous oscillation frequencies of the solution increase roughly as the inverse of the neutrino energy. Thus, it is expected that at low energies the integration should take longer times.
The results we report in Figure \ref{Fig-Suncpu} (solar neutrinos) 
stand for the relative error in the solution at the edge of the Sun for two energies, $E =1$ MeV and $10$ MeV, and confirm this point. This effect is more dramatic for DOPRI5 than for M4.
For fixed energy, M4 performs significantly better than DOPRI5.

Figure \ref{Fig-SNcpu} shows the results for SN neutrinos for $E=15$ MeV and $100$ MeV. 
The dependence of the cost as a function of $E$ follows the same pattern as with solar neutrinos.
 It is worth noticing the remarkable improvement achieved with the Magnus method in all cases in comparison 
 to the standard integration procedure.

\section{Discussion} \label{sec:discus}

For neutrinos propagating in a medium, numerical calculation of the flavor amplitudes 
is very often a costly process due to the highly oscillatory character of the solutions of 
the Schr\"odinger-like equation that governs the problem. Given the interest in analyzing 
the neutrino flavor evolution in a variety of settings, any novel procedure able to provide 
accurate approximations with a reduced computation time is of undeniable benefit. 
The purpose of this article is precisely to present one of such techniques, namely an 
explicit fourth-order integrator based on the ME for linear differential 
systems, that is equipped with an adaptive step-size strategy to render it more efficient 
in practice. 

To evaluate the performance of the new integrator M4 we compared it with the well known 
routine DOPRI5 based on an explicit 5th-order Runge\hs{.03cm}-Kutta method with variable step size. 
The comparison was made for three-mixed neutrinos evolving in the Sun and a type-II SN. 
We use realistic values for the oscillations parameters (mixing angles and square-mass differences). 
For simplicity, we modeled the matter densities in terms of simple analytical functions, 
but the procedure can be easily adapted to more complex situations, like a piecewise profile 
or one defined numerically.  In our numerical tests, we found that results with the same 
accuracy for the $\nu_e$ survival probability were produced by M4 with a computational 
cost between one and two orders of magnitude lower than DOPRI5, depending on the value 
of the neutrino energy. 

The applicability of the procedure is not constrained by the dimension of the system of linear 
differential equations. Therefore, it can be used to examine matter neutrino 
oscillations in the presence of one or more {\it sterile neutrinos}, i.e., massive singlet leptons 
that do not interact with the weak gauge bosons and mix with the standard active neutrinos.
As a result of the mixture of the active and sterile neutrinos, 
the PMNS matrix becomes a nonunitary part of a larger mixing matrix.  When looking
for possible effects of such non-unitarity in matter neutrino oscillations it might be advisable
to employ numerical methods that are not a affected by any intrinsic violation of unitarity.
As already emphazised, this is precisely an additional advantage of the Magnus integrator, 
which, by construction, will preserve the norm of the solution of the whole system, 
irrespective of the number of mixed neutrinos.

As has been formulated, our procedure cannot be used to integrate nonlinear differential 
equations like those arising in a SN core.  Nevertheless, a conveniently modified version 
might still be applied in these cases \cite{cas06}.

In Ref. \cite{montecarlo}, it has been presented a detailed proposal, based on a Monte Carlo method, for the (not preserving unitarity) numerical integration of the evolution operator for the two neutrino system. Although the aim of that work was not oriented to a systematic study of the CPU cost, the method seems to be quite expensive. The technique might be useful in long profiles, such as the core-collapse SN, but would not be the best choice in simpler situations, like the short, solar profile. 

In the case of two neutrino generations, a nonlinear transformation of the linear differential system
has been proposed as an alternative \cite{montecarlo}. In terms of hyperspherical polar coordinates, 
a new set of four nonlinear, real-valued equations is obtained. 
One of them, which stands for the norm of the solution, becomes a constant of motion and hence the norm is preserved, no matter the quality of the integration. The numerical integration proceeds actually with a system of three equations for which a preliminary analysis of singularities is needed.
An appropriate generalization of this scheme to three generations could be a further alternative starting point.
We have not yet explored this issue. 

In Perturbation Theory, the use of the interaction picture and the adiabatic basis are common tools. As far as the numerical integration of Eq. (\ref{eq:Hme}) is concerned, a change of picture transforms the slow varying coefficients system into a linear differential system whose coefficients oscillate rapidly, which actually is a harder problem. Anyhow, we carried out an implementation of such an approach \cite{iserlesHO} and convinced ourselves that any eventual advantage in the integration process does not compensate the extra algebra needed.  For this reason, we have not included it here.

Interrelation between Numerical Analysis and Physics has proved to be of mutual benefit. We hope that this work will help to motivate neutrino physics researchers to incorporate geometric ODE solvers as useful tools to do detailed numerical calculations on matter neutrino oscillations.

\begin{acknowledgments}

One of the authors (JCD) would like to thank the Departamento de 
F\'isica Te\'orica, Universidad de Valencia, where most of his work was done,
for its hospitality and its support via an EPLANET grant.
This work was supported in part by the Spanish MINECO under grants AYA2013-48623-C2-2 (JAO) and MTM2013-46553-C3 (FC), by DGAPA-UNAM grant PAPIIT IN112213 and
PASPA fellowship, and by CONACYT (M\'exico) grant  240666.
\end{acknowledgments}


\appendix*
\section{Computation of the matrix exponential} \label{sec:exp}

Given a Hermitian matrix $A$ of dimension three, we provide here
an efficient algebraic expression to explicitly compute the unitary matrix $\exp (itA)$,
where $t$ is a parameter. 
For caveats on the numerical computation of the exponential of a matrix see 
the classical paper by  Moler and Van Loan \cite{moler}.

Without loss of generality we will assume that $A$ is traceless, $\mathrm{Tr}(A)=0$. 
Were it not the case, then we can always factorize the problem: $\exp (i t A)= \exp (i t z  I) \exp(i t A_0)$, 
where $I$ stands for the identity matrix and $z = \mathrm{Tr}(A)/3$, $A_0=A-z I$.
Thus, the essential problem reduces to deal with a traceless 
matrix $A_0$. In practice, this conveys a number of important algebraic simplifications.

The eigenvalues of the (traceless) matrix $A_0$ are all real valued. They are given by the roots of the characteristic equation
\begin{equation} \label{eq:eigenlambdas}
\lambda^3-\frac{1}{2}\mathrm{Tr}(A_0^2)\lambda+\mathrm{Det}(A_0)=0.
\end{equation}

The explicit solutions may then be written down as
\begin{equation}
\lambda_k=\pm 2\sqrt{\frac{p}{3}}\cos\left[ \frac{1}{3}\arccos\left(\frac{3q}{2p}
\sqrt{\frac{p}{3}}\right)-\frac{2\pi k}{3}\right] ,
\end{equation}
for $(k=0,1,2)$, where $p\equiv \mathrm{Tr}(A_0^2)/2$, and $q\equiv \mathrm{Det}(A_0)$. The positive sign corresponds to
$q\le 0$, otherwise the negative sign applies.
The fact $p>0$ ensures the real character of the roots. 

Finally, using Putzer's algorithm \cite{Putzer} one gets
\begin{eqnarray}\label{eq:Putzer}
\exp(itA_0)&=& \exp(i\lambda_0 t)[(1-\lambda_0 (r_0-\lambda_1 r_1))I \nonumber\\
& &+(r_0+\lambda_2 r_1)A_0+r_1A_0^2],
\end{eqnarray}
where the eigenvalues have been relabeled such that now $\lambda_0<\lambda_1<\lambda_2$, and 
\begin{eqnarray}\label{eq:r-s}
r_0&=&-\frac{1-\exp(iat)}{a}, \\
r_1&=&-\frac{1}{a-b}\left(\frac{1-\exp(iat)}{a}-
\frac{1-\exp(ibt)}{b}\right), \nonumber
\end{eqnarray}
with $a=\lambda_1-\lambda_0$, $b=\lambda_2-\lambda_0$. The formulae (\ref{eq:r-s}) for the complex coefficients $r_0,r_1$, cope with eventual situations where $a,b\ll 1$. 

The most expensive part of the computation comes from the matrix product $A_0^2$.
Thus, considerable amount of CPU time is saved if the explicit numerical matrix computation $A_0^2$ 
in (\ref{eq:Putzer}) is avoided and, instead, only the matrix--vector product 
$A_0\Psi(t)$ and $A_0(A_0\Psi(t))$ are evaluated on the solution vector $\Psi$ along the integration.
Notice that in (\ref{eq:eigenlambdas}) the computation of the whole matrix $A_0^2$ can also be avoided.

\end{document}